\documentclass{PoS}
\usepackage{setspace}
\usepackage[TABBOTCAP]{subfigure}
\newlength{\figlen}
\newlength{\newfiglen}
\title{Development of the photomultiplier tube readout system for the first Large-Sized Telescope of the Cherenkov Telescope Array}

\ShortTitle{Development of the PMT readout system for the first LST of the CTA}

\author{{\fontsize{10.3pt}{0cm}\selectfont \speaker{Shu Masuda}$^{a}$,  Yusuke Konno$^{a}$, 
Juan Abel Barrio$^{b}$, 
Oscar Blanch Bigas$^{c}$, 
Carlos Delgado$^{d}$, 
Llu\'is Freixas Coromina$^{d}$, 
Shuichi Gunji$^{e}$, 
Daniela Hadasch$^{f}$, 
Kenichiro Hatanaka$^{a}$, 
Masahiro Ikeno$^{g}$,  
Jose Maria Illa Laguna$^{c}$, 
Yusuke Inome$^{h}$,  
Kazuma Ishio$^{f}$,  
Hideaki Katagiri$^{i}$, 
Hidetoshi Kubo$^{a}$, 
Gustavo Mart\'inez$^{d}$, 
Daniel Mazin$^{f}$, 
Daisuke Nakajima$^{f}$, 
Takeshi Nakamori$^{e}$, 
Hideyuki Ohoka$^{f}$, 
Riccardo Paoletti$^{j}$, 
Stefan Ritt$^{k}$,  
Andrea Rugliancich$^{j}$, 
Takayuki Saito$^{a}$, 
Karl-Heinz Sulanke$^{l}$,  
Junki Takeda$^{e}$, 
Manobu Tanaka$^{g}$, 
Shunsuke Tanigawa$^{a}$, 
Luis \'Angel Tejedor$^{b}$, 
Masahiro Teshima$^{f,m}$, 
Yugo Tsuchiya$^{a}$, 
Tomohisa Uchida$^{g}$, 
Tokonatsu Yamamoto$^{h}$ 
and the LST team for the CTA Consortium}\\%\footnote{Full consortium author list at http://cta-observatory.org}\\
\begin{spacing}{0.8}
{\scriptsize   
$^{a}$Kyoto University, Sakyo, Kyoto 606-8502, Japan; 
$^{b}$Universidad Complutense de Madrid, Av. Complutense s/n, 28040 Madrid, Spain; 
$^{c}$Institut de F\'isica d'Altes Energies, Edifici Cn, Campus UAB, 08193 Bellaterra, Spain; 
$^{d}$CIEMAT, Avda. Complutense 22, 28040 Madrid, Spain; 
$^{e}$Yamagata University, Yamagata 990-8560, Japan; 
$^{f}$Institute for Cosmic Ray Research, University of Tokyo, Kashiwa, Chiba 277-8582, Japan; 
$^{g}$High Energy Accelerator Research Organization (KEK), Tsukuba, Ibaraki 305-0801, Japan; 
$^{h}$Konan University, Kobe, Hyogo, 658-8501 Japan; 
$^{i}$Ibaraki University, Mito, Ibaraki 310-8512, Japan; 
$^{j}$Universita degli Studi di Siena \& INFN, Pisa, I-53100 Siena, Italy; 
$^{k}$Paul Scherrer Institute, 5232 Villigen, Switzerland; 
$^{l}$Deutsches Elektronen-Synchrotron, Platanenallee 6, D-15738 Zeuthen, Germany; 
$^{m}$Max-Planck-Institute f\"ur Physik, F\"ohringer Ring 6, D-80805 M\"unchen, Germany\\
E-mail: \email{masuda@cr.scphys.kyoto-u.ac.jp}, \email{konno@cr.scphys.kyoto-u.ac.jp}%\\
}
\end{spacing}
}

\abstract{{\small The Cherenkov Telescope Array (CTA) is the next generation ground-based very high energy gamma-ray observatory. The Large-Sized Telescope (LST) of CTA targets 20 GeV -- 1 TeV gamma rays and has 1855 photomultiplier tubes (PMTs) installed in the focal plane camera. With the 23 m mirror dish, the night sky background (NSB) rate amounts to several hundreds MHz per pixel. In order to record clean images of gamma-ray showers with minimal NSB contamination, a fast sampling of the signal waveform is required so that the signal integration time can be as short as the Cherenkov light flash duration (a few ns). We have developed a readout board which samples waveforms of seven PMTs per board at a GHz rate. Since a GHz FADC has a high power consumption, leading to large heat dissipation, we adopted the analog memory ASIC ``DRS4''. The sampler has 1024 capacitors per channel and can sample the waveform at a GHz rate. Four channels of a chip are cascaded to obtain deeper sampling depth with 4096 capacitors. After a trigger is generated in a mezzanine on the board, the waveform stored in the capacitor array is subsequently digitized with a low speed (33 MHz) ADC and transferred via the FPGA-based Gigabit Ethernet to a data acquisition system. Both a low power consumption (2.64 W per channel) and high speed sampling with a bandwidth of  $>$300 MHz have been achieved. In addition, in order to increase the dynamic range of the readout we adopted a two gain system achieving from 0.2 up to 2000 photoelectrons in total. We finalized the board design for the first LST and proceeded to mass production. Performance of produced boards are being checked with a series of quality control (QC) tests. We report the readout board specifications and QC results.}

}

\FullConference{\begin{spacing}{0.9}The 34th International Cosmic Ray Conference,\\
		30 July- 6 August, 2015\\
		The Hague, The Netherlands\end{spacing}}

\begin{document}

\section{Introduction}
Very high energy (VHE) cosmic gamma rays provide clues of understanding the origin of cosmic rays, the mechanism of particle acceleration around black holes and the nature of cold dark matter. The imaging atmospheric Cherenkov telescopes (IACTs) are nowadays the standard detectors for VHE gamma rays. Multiple telescopes take a Cherenkov-light stereoscopic image of an extensive air shower (EAS) induced by a gamma ray entering the atmosphere. The Cherenkov Telescope Array (CTA) is the next generation VHE gamma-ray observatory. The Large-Sized Telescope (LST)~\cite{ICRC2015LST} has a 23 m mirror dish and targets 20 GeV -- 1 TeV gamma rays. Its camera has 1855 photomultiplier tubes (PMTs) as photon detectors. With its large mirror dish, the night sky background (NSB) photons are also detected at a rate of  several hundreds MHz per pixel. On the other hand, the duration of the Cherenkov light flash is only a few nanoseconds. In order to suppress NSB contamination to shower images and to improve the signal-to-noise ratio, it is necessary to sample the signal waveform at rate of $\sim$1 GHz and to reduce the signal integration time down to a few ns. 

However, a GHz flash analog-to-digital converter (FADC) is in general expensive and has a high power consumption with large heat dissipation. On the other hand, the ``DRS4'' chip~\cite{Ritt2010486}, which is an application specific integrated circuit (ASIC) of switched capacitor arrays (SCAs), can sample at a rate of GHz with low power consumption. We have developed a readout board adopting the DRS4 and have demonstrated that it can properly digitize the fast PMT signal. We report the design and the performance of the final version of the readout board to be installed in the camera of the first LST.

\section{Design of the readout system}
\subsection{Overview} %~\cite{ICRC2015PMT} 
\begin{figure}[hbtp]
\centering
\includegraphics[]{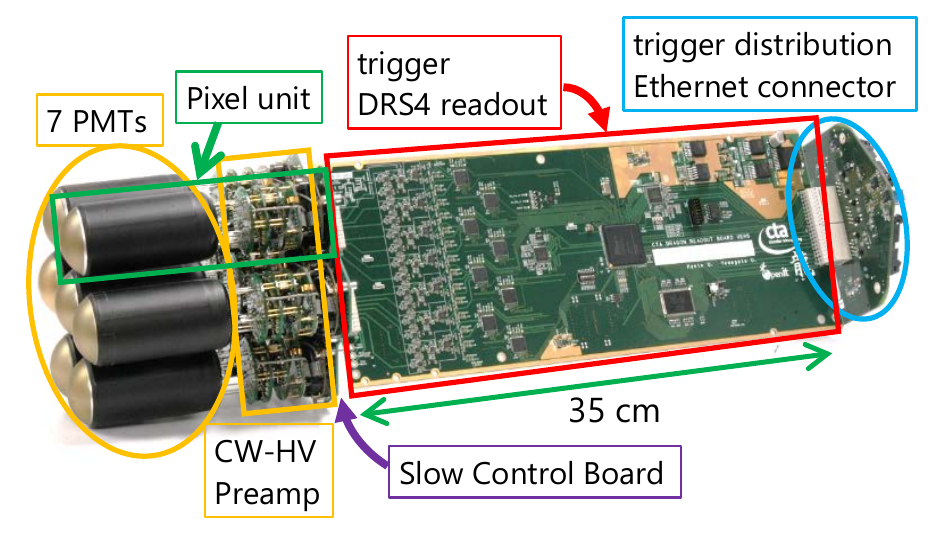}
\caption{Photograph of a 7-PMT module to be installed in the camera of the first LST.}
\label{PMTcluster}
\end{figure}

The focal plane camera of the LST has 1855 PMT pixels and consists of 265 modules with seven pixel units each.  Figure~\ref{PMTcluster} shows a photograph of a 7-PMT module to be installed in the camera of the first LST. The 7-PMT module consists also of a slow control board (SCB)~\cite{ICRC2015SCB}, a DRS4 readout board and a backplane board which has several connectors for the $+$24 V power supply and for interfaces with other external modules. Each pixel unit is composed of a PMT, a Cockcroft-Walton high voltage supply circuit and a preamplifier ASIC named PACTA. 
\begin{figure}[hbtp]
\centering
\includegraphics[]{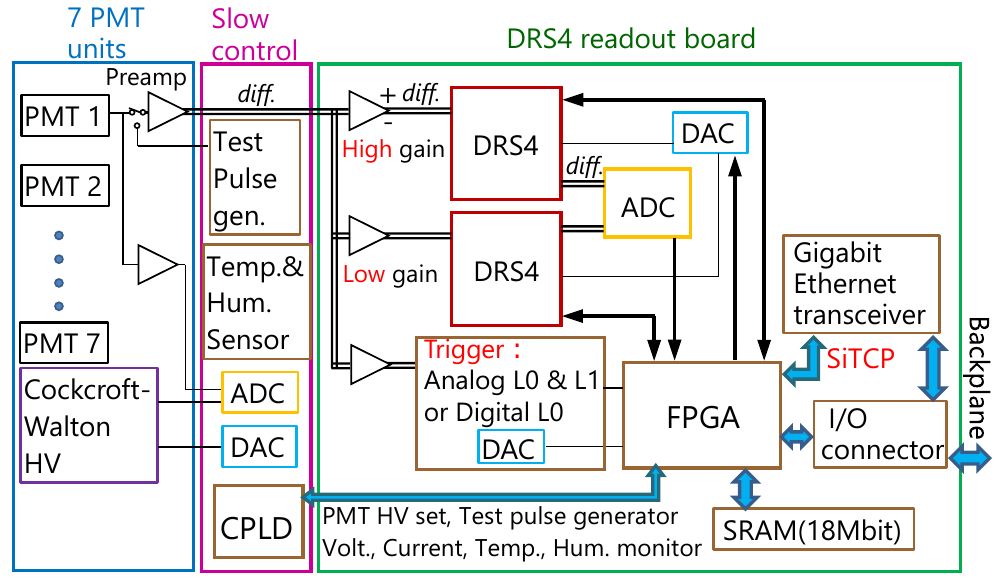}
\caption{Block diagram of a 7-PMT camera module.}
\label{block}
\end{figure}

Figure~\ref{block} shows the block diagram of the readout system. On the DRS4 readout board, each signal from the PACTA preamplifiers is divided into three lines: a high gain line, a low gain line and a trigger line. The high gain and low gain signals are sampled by the DRS4 chip at a 1 GHz rate. When a trigger signal is generated, the stored waveform is digitized by an external slow ADC at a rate of $\sim$33 MHz. The digitized data is then sent to a field programmable gate array (FPGA) and transferred to the data storage server through the backplane board and switches via the Gigabit Ethernet transceiver. This technology of transmission of waveform data and slow control is available by implementing SiTCP~\cite{Uchida08} on FPGA. The Xilinx Spartan-6 FPGA controls eight DRS4 chips, a 12-bit ADC, a static random access memory (SRAM) which can store 18 Mbits of data and a digital-to-analog converter (DAC) used for correction of the input and output ranges of DRS4s. In addition, it communicates with a complex programmable logic device (CPLD) on the SCB and a FPGA on the backplane board via serial peripheral interface (SPI).

\begin{figure}[hbtp]
\centering
\includegraphics[]{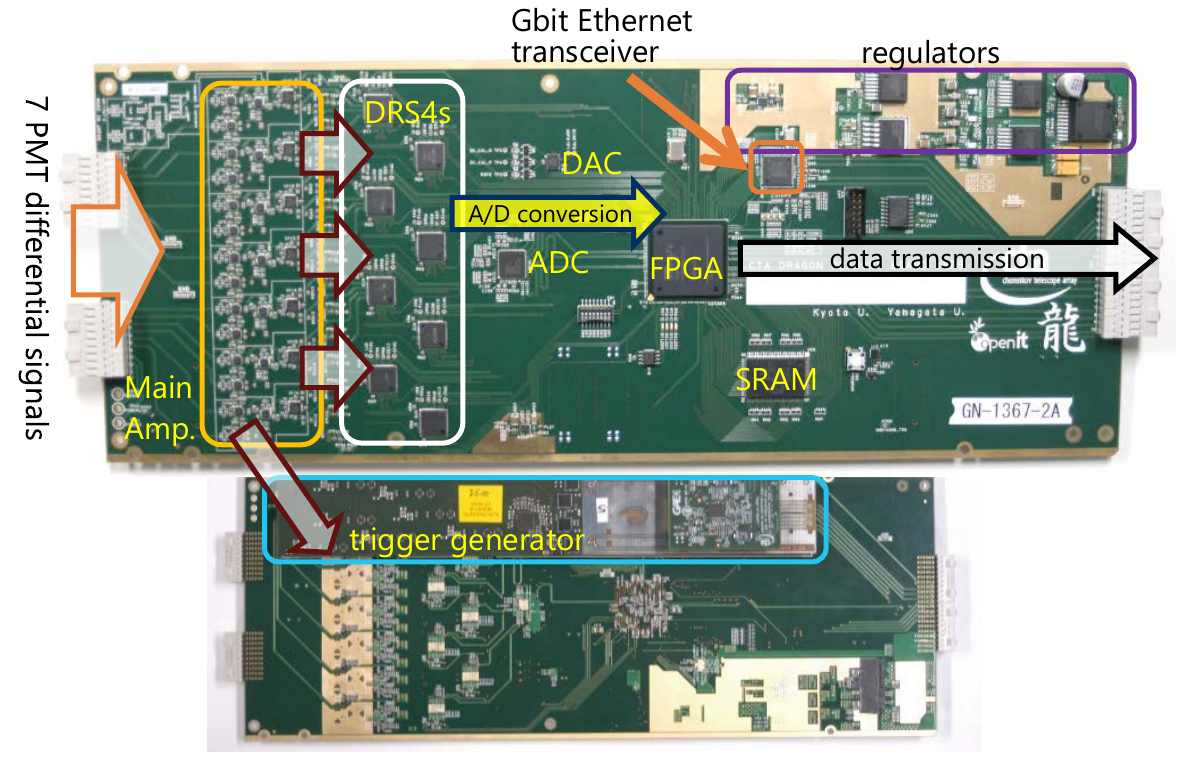}
\caption{Photograph of front side (top) and back side (bottom) of the DRS4 readout board.}
\label{Dragon}
\end{figure}

Figure~\ref{Dragon} shows a photograph of DRS4 readout board with a size of 35 cm $\times$ 13.6 cm. A SCB and a backplane board are connected to the left and right side connectors respectively. The power consumption per channel without pixel units is 2.64 W. It has four connectors as the interface to mezzanine boards of trigger generator circuit. Details of the parts on the board are described in the following subsections. 

\subsection{DRS4}
The Domino Ring Sampler (DRS) has been developed at the Paul Scherrer Institute (PSI), Switzerland, for the MEG experiment~\cite{MEG13}. The latest version, DRS4, is also used in the MAGIC experiment~\cite{Aleksic2015}. The DRS4 chip has nine differential input channels with 950 MHz bandwidth and each channel has 1024 storage capacitors whose write switches are operated via a chain of inverters named domino wave circuit. The sampling speed can be changed from 700 MHz up to 5 GHz according to reference clock from the FPGA. In addition, a low noise level of 0.35 mV (rms) (after offset correction) and a low power consumption of 17.5 mW per channel at 2 GHz sampling are demonstrated~\cite{DRS4datasheet}. Eight DRS4 chips are loaded on the readout board and each chip is connected to two PMT channels. It is possible to cascade up to eight DRS4 channels for a deeper sampling depth. Cascading is realized in the following way; there is an 8-bit register in the chip, each bit of which corresponds to one DRS4 channel. A channel is activated only when the corresponding bit of the register is high. Connecting the same signal line to multiple DRS4 channels and activating them one after another, the sampling depth can be summed. For the readout board, four DRS4 channels are cascaded, which means the signal of each input is sampled in 4096 capacitors and the sampling depth is $\sim$4 $\mu$s in the case of 1 GHz sampling.

\subsection{Main amplifier}
\begin{figure}[hbtp]
\centering
\includegraphics[width=0.8\linewidth]{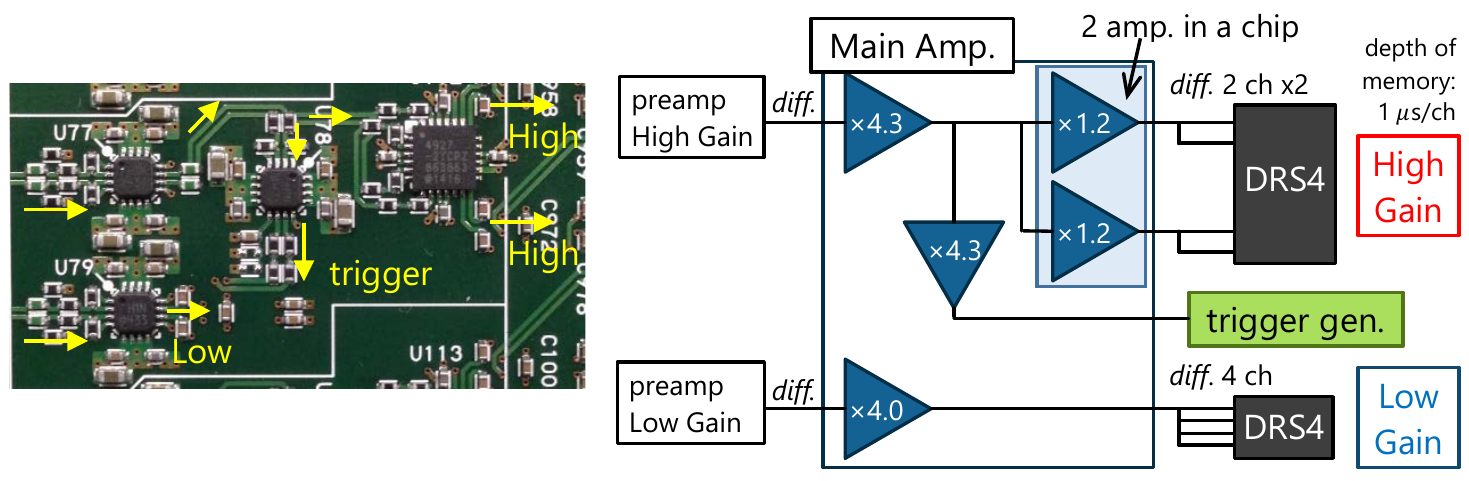}
\caption{Photograph of the main PMT channel amplifier (left) and block diagram showing its interfaces (right).}
\label{mainamp}
\end{figure}

Figure~\ref{mainamp} shows a photograph and a block diagram of the main amplifier on the DRS4 readout board. It is designed to achieve wide bandwidth of $>$300 MHz and to have a dynamic range up to 2000 photoelectrons by using a high speed current feedback amplifier, Analog Devices ADA4927. Five amplifiers are used per PMT channel and two of them are integrated in one chip. As described above, in the readout board, four DRS4 channels are cascaded and thus, four identical signal lines per channel must be prepared. It was known from simulation that the bandwidth can be significantly improved by increasing the number of amplifiers dealing with the four signal lines because the capacitance load per amplifier is decreased. Therefore, the main amplifier of high gain PMT channels is designed so that four DRS4 channels are fed by two amplifiers.

\subsection{Trigger mezzanines}
There are two types of trigger logic; analog trigger and digital trigger. For the analog trigger%(e.g.~\cite{trigger})
, level 0 and 1 (L0, L1) mezzanine boards for trigger generation are mounted on the DRS4 readout board as shown at the bottom of Figure~\ref{Dragon}. The trigger circuit on the L0 mezzanine generates an analog waveform which is the sum of pulses from seven PMT channels. This pulse is distributed to the L1 mezzanines on neighbor modules by a backplane board. On each L1 mezzanines, the pulses from a backplane board are summed up and discriminated. By this method, the energy threshold can be decreased. On the other hand, for the digital trigger, only a L0 mezzanine board is mounted. PMT signals are discriminated pixel by pixel and backplane boards make the decision to trigger by a geometrical pixel hit pattern. %It is possible to choose the both trigger types by replacing the mezzanines. 
The readout board is compatible with both trigger types, which can be exchanged by simply replacing the mezzanines.

%\subsection{FPGA-based Gigabit Ethernet (SiTCP)}
%Transmission of waveform data and slow control are performed via Ethernet using two devices: FPGA and Gigabit Ethernet transceiver (PHY). This technology is available by implementing SiTCP~\cite{Uchida08} on FPGA. The SiTCP circuit costs only $\sim$3000 Slices (where a Slice is a minimal logic block in the Xilinx FPGA), which is small enough to implement user circuits on the same FPGA.

\section{Performance of the readout system}
\subsection{Fundamental performance}
We have evaluated the performance of the DRS4 readout board and several results of measurements are shown below.

\begin{figure}[hbtp]
\centering
\includegraphics[width=\newfiglen]{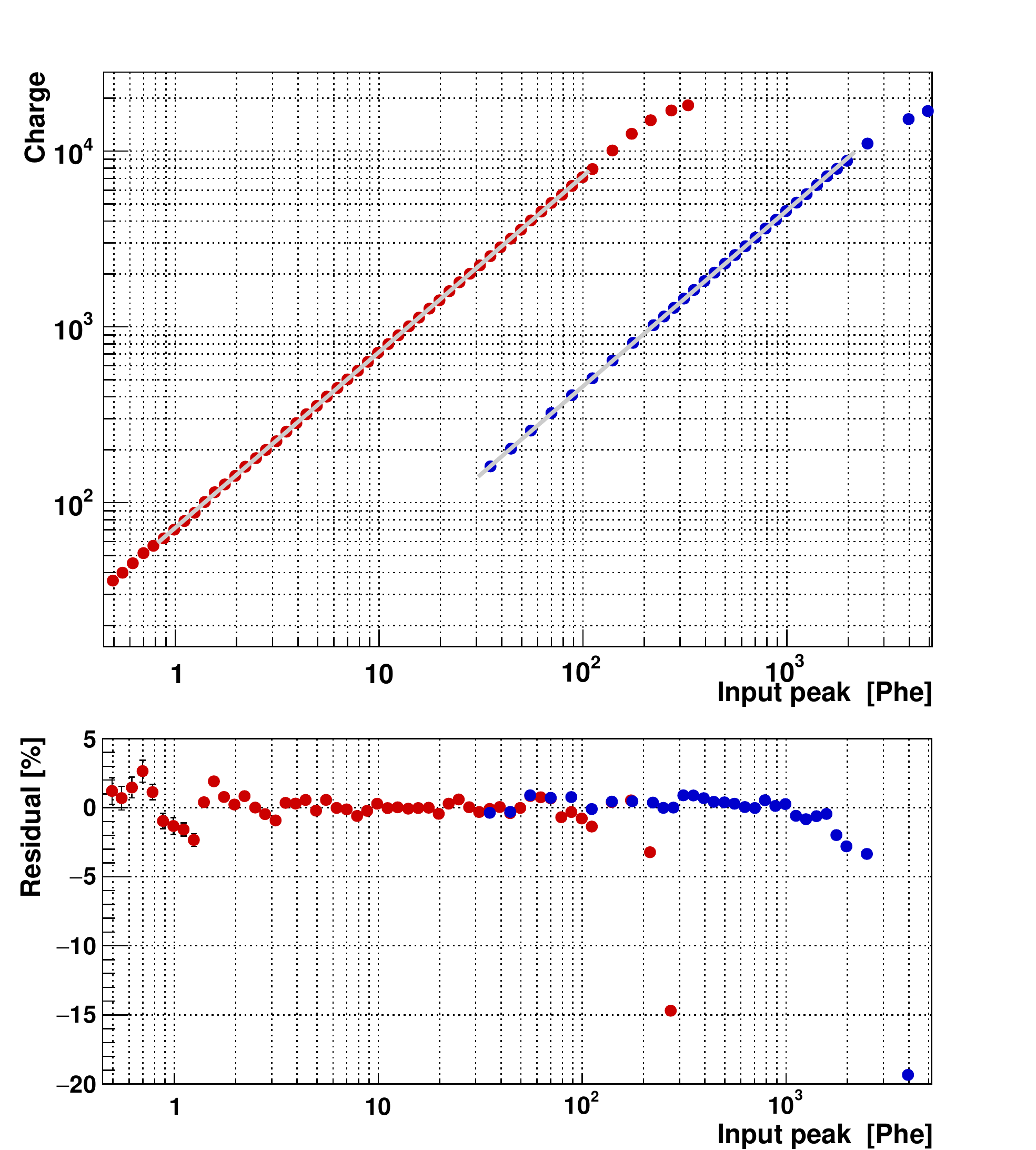}
\caption{Dynamic ranges of high (red) and low (blue) gain lines of the DRS4 readout board (top) and residuals between measured values and the fit line (bottom).}
\label{drange}
\end{figure}

\begin{figure}[hbtp]
\centering
\includegraphics[width=0.98\newfiglen]{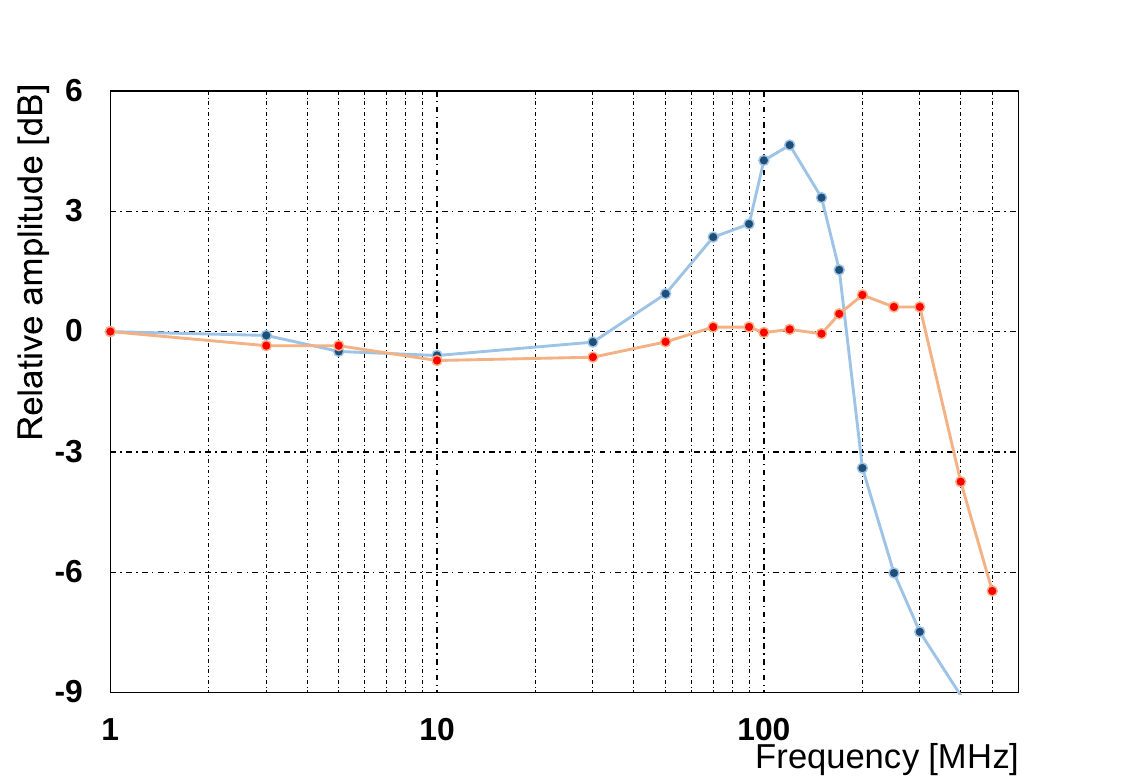}
\caption{Bandwidths of high (red) and low (blue) gain line.}
\label{bandwidth}
\end{figure}

First, we evaluated the main amplifier. Dynamic ranges of high-gain and low-gain lines are shown in Figure~\ref{drange}. It is demonstrated that linearity within 5 \% is totally guaranteed from below 1 up to $>$2000 photoelectrons. Figure~\ref{bandwidth} shows the bandwidth of the high gain and low gain lines. The high-gain line achieves a bandwidth of $>$300 MHz, which is wide enough to measure Gaussian like waveforms with a full width at half maximum (FWHM) of as fast as 2 ns. %The low gain one has only $\sim$180 MHz. It is not a big problem because low gain channels deal with signals with more than 30 photoelectrons, which is far larger than NSB signals.
On the other hand, the low gain line achieves $\sim$180 MHz. Since this channel processes signals significantly larger (more than 30 photoelectrons) than typical NSB, this bandwidth limitation is not considered a relevant shortcoming.

\begin{figure}[hbtp]
\centering
\includegraphics[width=\figlen]{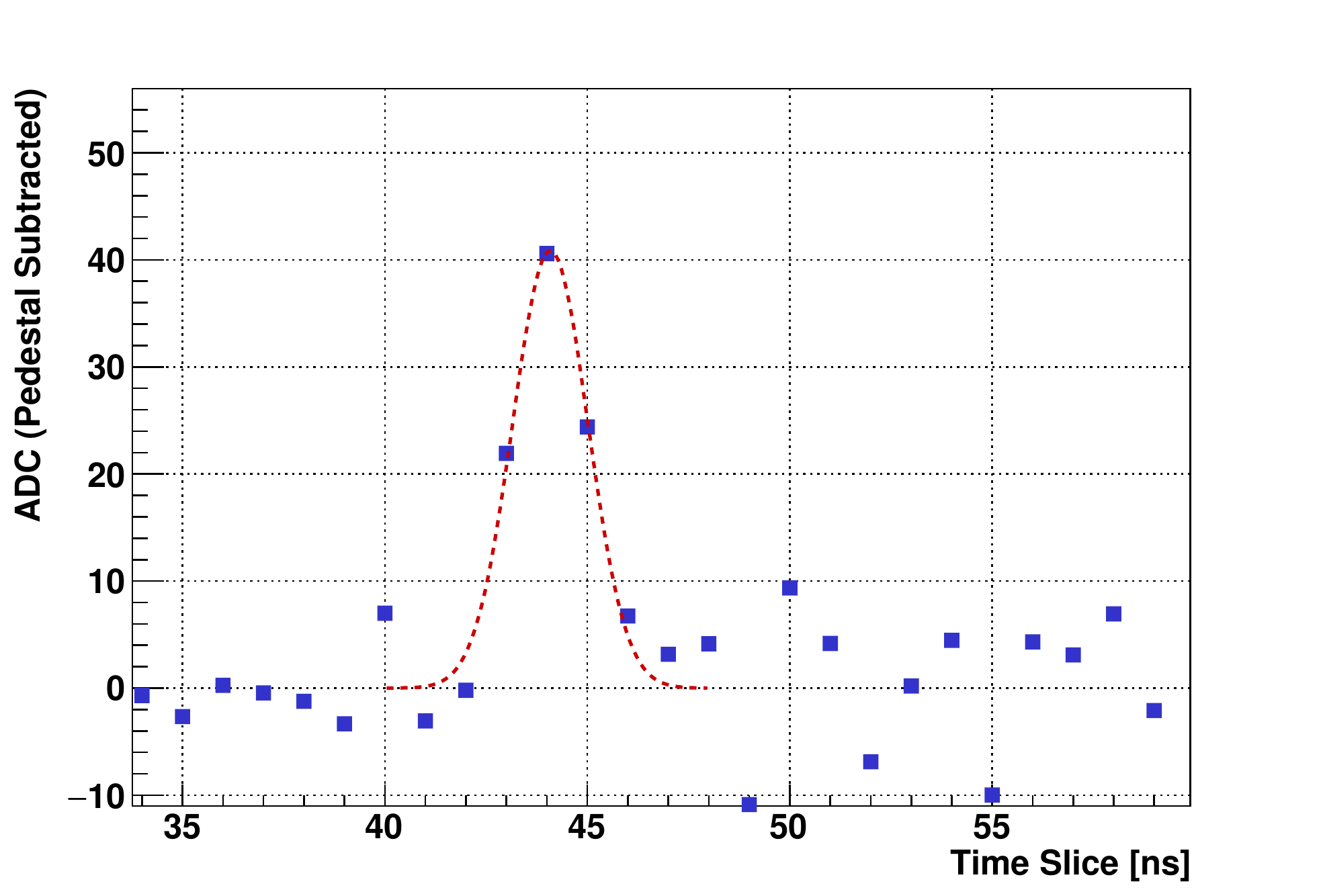}
\caption{Pulse waveform of a PMT single photon signal sampled by the DRS4 readout board at a rate of 1 GHz.}
\label{sp}
\end{figure}

\begin{figure}[hbtp]
\centering
\includegraphics[width=\figlen]{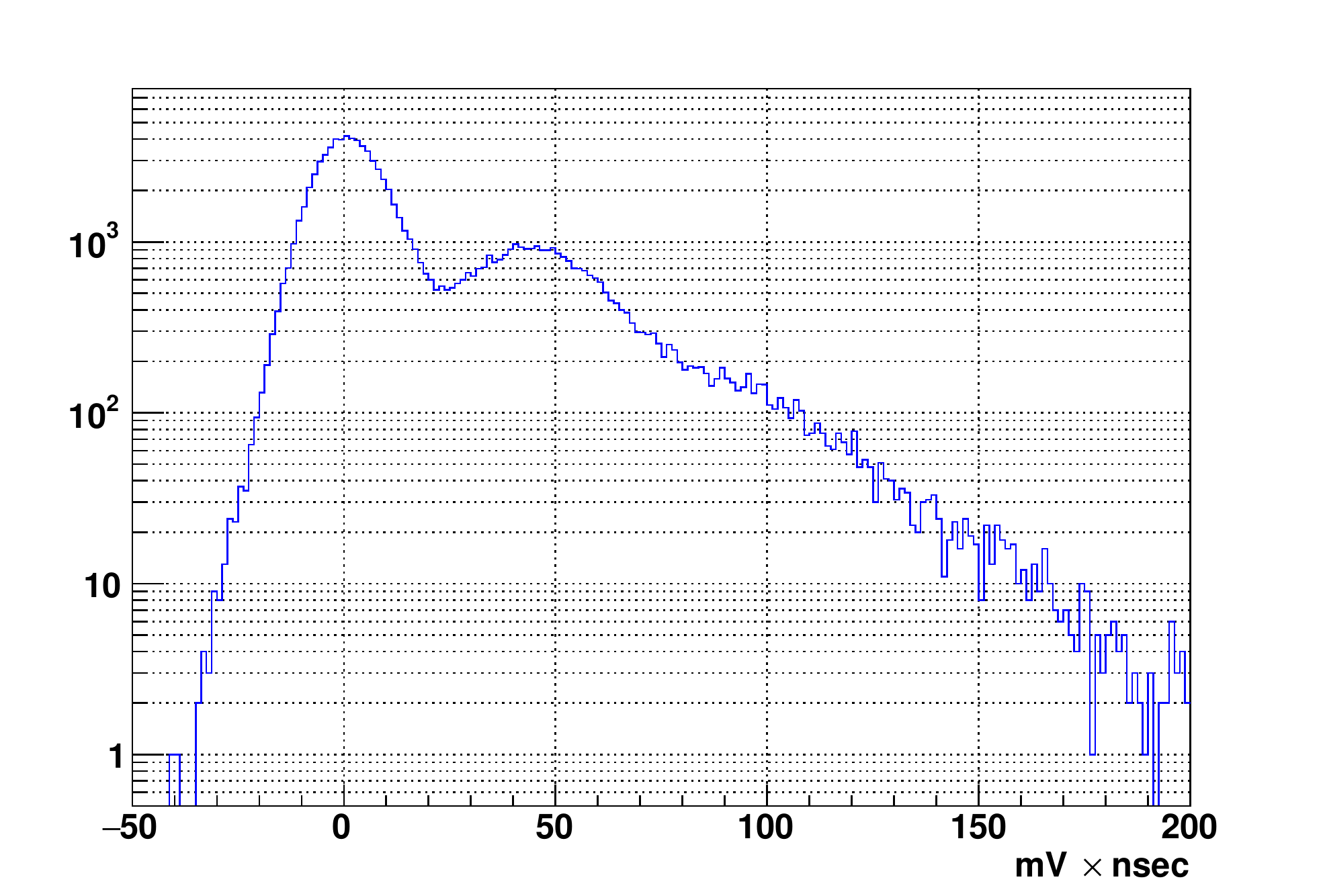}
\caption{Charge distribution of single photoelectrons. The horizontal axis is in units of mV $\times$ ns.}
\label{spe}
\end{figure}

We also performed measurements of signals from a PMT via the SCB. Figure~\ref{sp} shows a pulse waveform of a single photon signal from a PMT sampled by the DRS4 readout board at a rate of 1 GHz. It was measured with weak light from a LED and with a PMT gain of 40000. As shown in the figure, we successfully digitized the PMT signal with a FWHM of 2--3 ns. We analyzed the waveform data and created a single photoelectron charge distribution (Fig.~\ref{spe}). As shown in the figure, the single photoelectron peak is clearly seen. Both the pedestal peak and single photoelectron peak are fit by Gaussian distributions, and the signal-to-noise ratio, defined as (single-photoelectron mean -- pedestal mean) / pedestal rms, is calculated to 5.26. In addition we also measured the time jitter of test pulses from a trigger and determined that it is lower than $\sim$1 ns (rms) at a sampling rate of 1 GHz.

\subsection{Quality control results}
We have started to mass-produce $\sim$300 readout boards in total for the first LST. %So far 69 boards are produced and gone the quality control tests. 
69 boards have been produced by now and submitted to the LST quality control process. Figure~\ref{Noise} shows a distribution of amplitude rms of zero-signal events. As shown in these figures, the noise level of all boards is less than 0.2 photoelectrons. 
%Figure~\ref{Noise} shows a distribution of noise level. The left of the figure shows a distribution of means of sampling points' rms of none-signal events, and right one shows a distribution of integral values of three neighboring sampling points. As shown in these figures, noise level of all boards are less than 0.2 photoelectrons against the both two methods. Figure~\ref{CT} shows a distribution of maximum crosstalk level of each channel. Crosstalk levels of all channels of all boards are less than 0.6 \%.
%Two different procedures were used to qualify the board for low noise operation [explain both]. Figure~\ref{Noise} shows the distribution of noise levels after testing the 69 boards with both procedures. It can be seen that the noise level of all tested boards is well below 0.2 photoelectrons, no matter which method is used to define this specification. 
Figure~\ref{CT} shows the distribution of the maximum crosstalk levels for the tested channels, which are measured by the test pulses injected from the SCB. All of them showed levels below 0.6 \%. In addition, non-linearity of all boards is less than $\pm$5 \%.
\if0\begin{figure}[hbtp]
\centering
%\subfigure{\includegraphics[width=0.4\linewidth]{NoiseRMS_HG}}%
%\hspace{3cm}%
%\subfigure{\includegraphics[width=0.4\linewidth]{CoherentNoseSearched_HG}}
\includegraphics[width=0.4\linewidth]{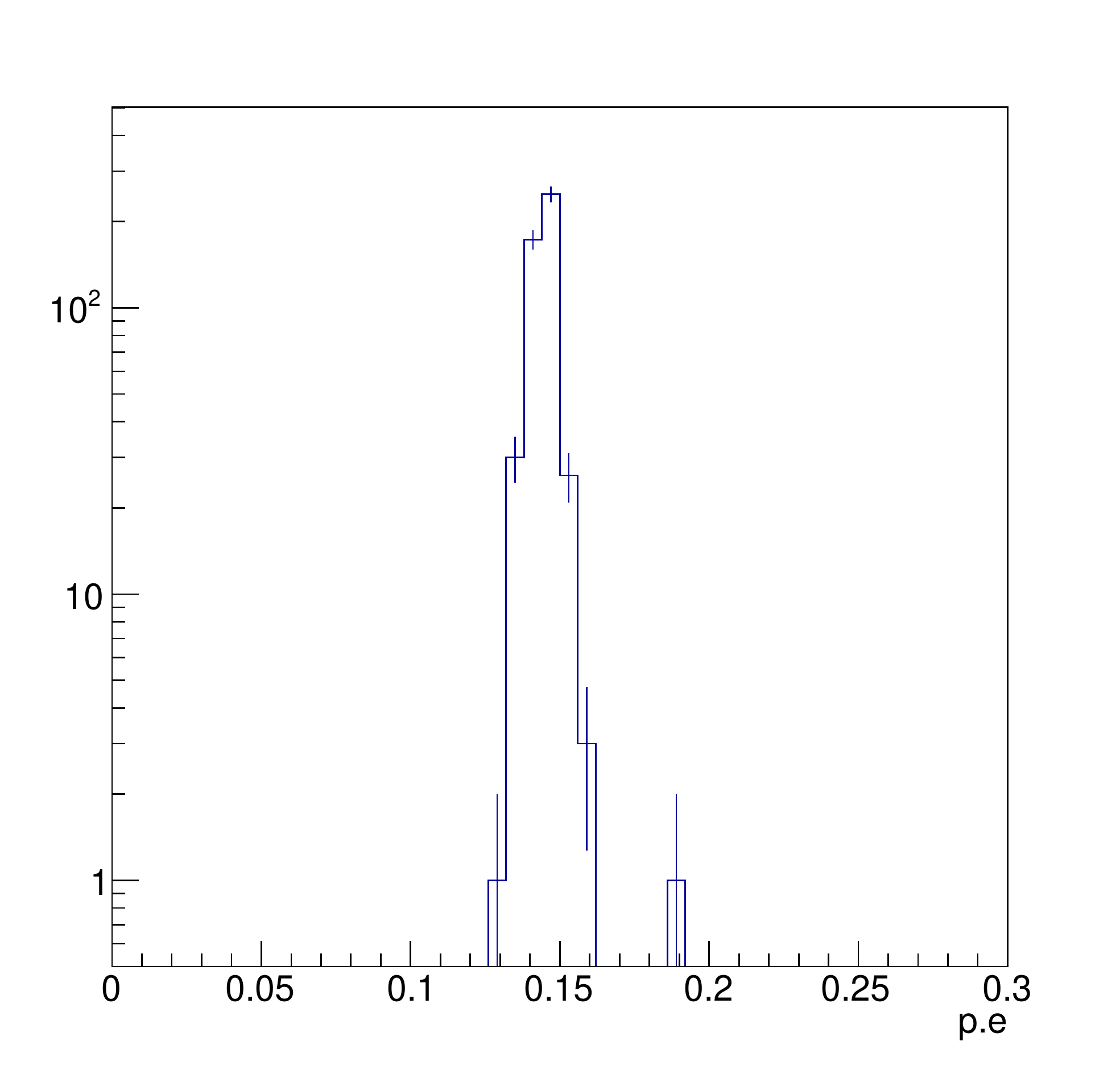}
\caption{Distribution of means of noise rms for all channels of 69 quality checked boards. The horizontal axis is in units of photoelectron.}
\label{Noise}
\end{figure}
\begin{figure}[hbtp]
\centering
\includegraphics[width=0.4\linewidth]{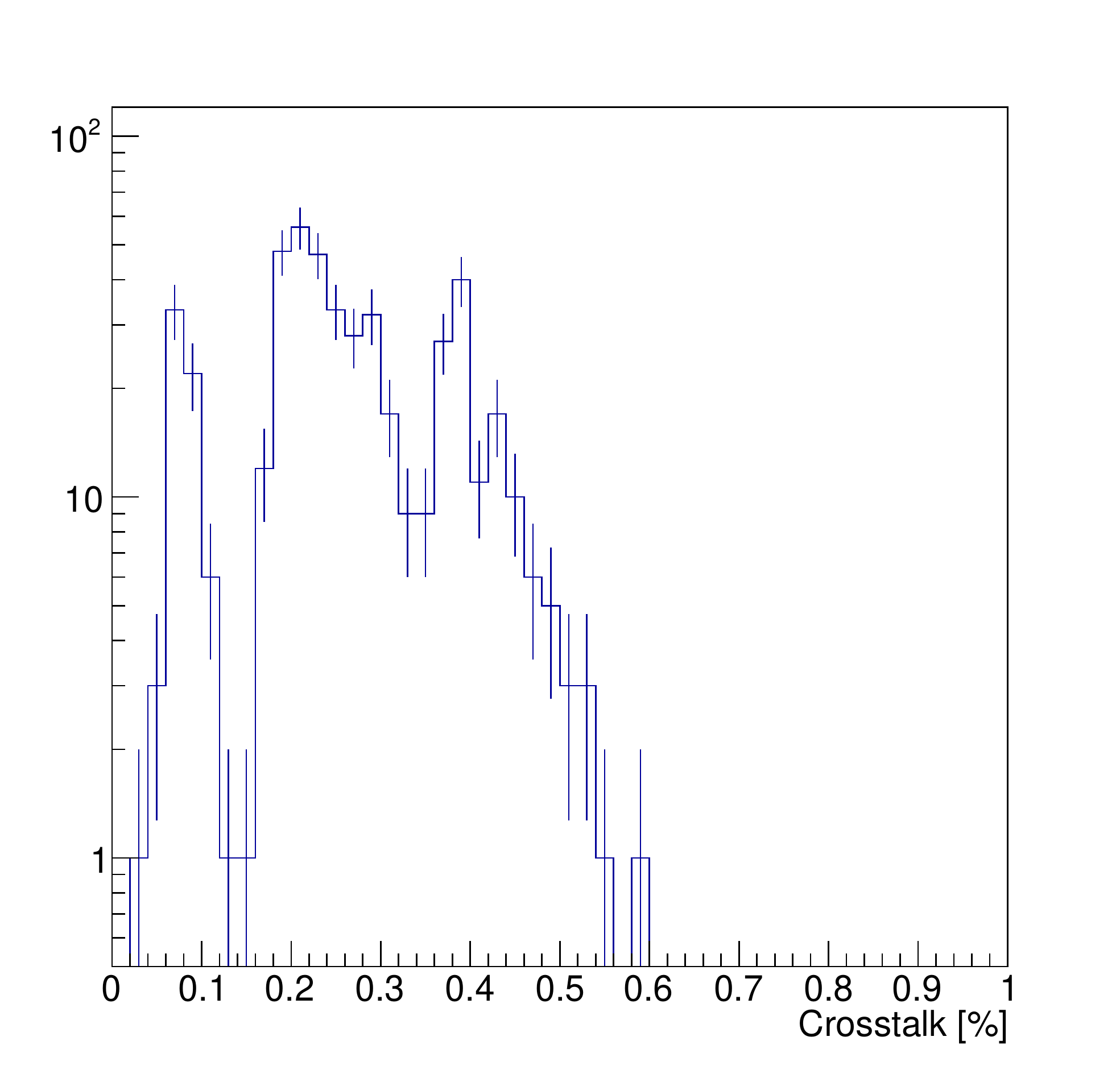}
\caption{Distribution of biggest crosstalk levels for all channels of 69 quality checked boards. The horizontal axis is in units of percentage to input amplitude of neighbor channels.}
\label{CT}
\end{figure}\fi
\begin{figure}[hbtp]
\begin{minipage}[t]{0.48\columnwidth}
\centering
\includegraphics[width=0.8\linewidth]{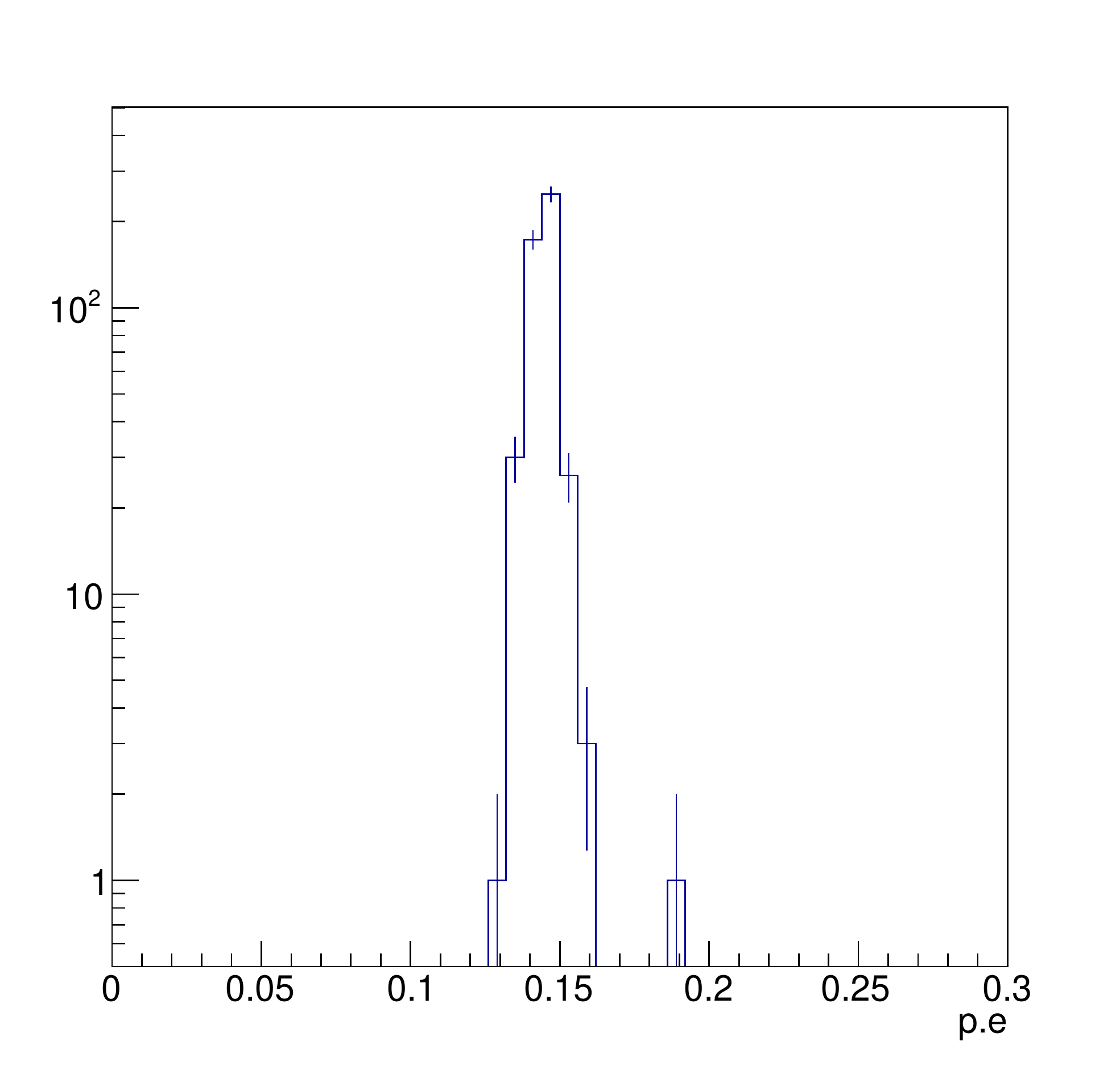}
\caption{Distribution of noise for all channels of 69 quality checked boards, which is defined as amplitude rms of zero-signal events. The horizontal axis is in units of photoelectron.}
\label{Noise}
\end{minipage}
\hspace{0.03\columnwidth}
\begin{minipage}[t]{0.48\columnwidth}
\centering
\includegraphics[width=0.8\linewidth]{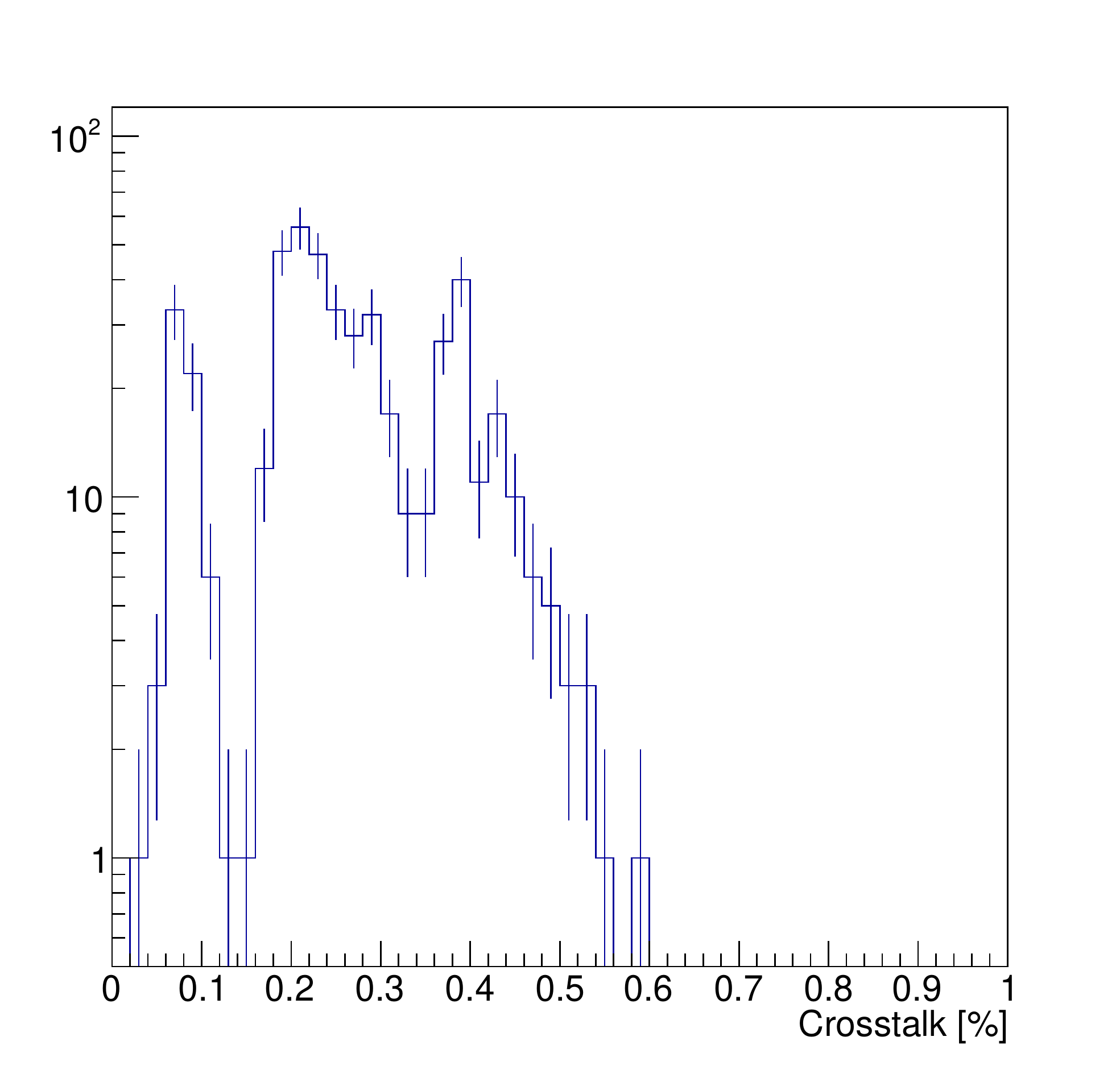}
\caption{Distribution of biggest crosstalk levels for all channels of 69 quality checked boards. The horizontal axis is in units of percentage to input amplitude of neighbor channels.}
\label{CT}
\end{minipage}
\end{figure}

\section{Conclusion}
We have developed and completed a readout system to be installed in the first Large-Sized Telescope of CTA. We demonstrated that it can successfully sample waveforms of a PMT by using the analog memory ASIC DRS4 chips at a 1 GHz rate. It achieves a low power consumption of 2.64 W/ch, a wide bandwidth of $>$300 MHz at high gain line, and a wide dynamic range from below 1 up to 2000 photoelectrons.

\acknowledgments
We gratefully acknowledge support from the agencies and organizations listed under Funding Agencies at this website: http://www.cta-observatory.org/. We also acknowledge the great help from Open-It Consortium on the hardware development of the DRS4 readout board.

\bibliographystyle{JHEP2}
\bibliography{myref}

\end{document}